\documentclass[12pt]{article}
\usepackage{graphicx,amsmath,amssymb,amsfonts,color}
\usepackage{citesort,epsfig}

\input{text/alphaseries.pre}
\input{text/alphaseries.fig}

\begin{document}


\begin{titlepage}

\begin{tabular}{l}
\noindent\DATE
\end{tabular}
\hfill
\begin{tabular}{l}
\PPrtNo
\end{tabular}

\vspace{1cm}

\begin{center}
\renewcommand{\thefootnote}{\fnsymbol{footnote}}
{
\Large \TITLE
}

\vspace{1.25cm}
{\large  \AUTHORS}

\vspace{1.25cm}

\INST
\end{center}

\vfill

\ABSTRACT                 

\vfill

\end{titlepage}

\tableofcontents 

\section{Introduction}

\label{sec:introduction}

The theory of Quantum Chromodynamics (QCD) depends on the fundamental gauge
coupling strength $\alpha_{s}$ and suitably defined quark mass
parameters $m_{f}$.
For applications to hard scattering processes with hadrons
in the initial state,
we also need the universal parton distribution functions (PDFs)
that characterize the partonic structure of those hadrons.
These PDFs are determined by global QCD analysis,
using input from a variety of well-established
experimental measurements.

In the standard CTEQ analysis \cite{cteq6}, the focus is on determining 
the parton distribution functions.  The value of $\alpha_{s}(m_{Z})$ is fixed 
at the world average value, which is dominantly based on dedicated 
measurements such as those from LEP, where it was determined in 
processes that are free of the complications of hadronic structure.

However, the interplay between the coupling strength $\alpha_{s}$ and
strong dynamics is an interesting subject in itself.
Many attempts have been made to extract $\alpha_{s}(m_{Z})$ from individual 
experiments at the HERA $\mathrm{ep}$ collider, at the Tevatron 
$\mathrm{p\bar{]}}$ collider, and from combined analyses of
several hadronic experiments.
For such studies, as well as to assess the additional uncertainty in 
predictions caused by the uncertainty in $\alpha_{s}(m_{Z})$,
it is important to have PDF sets available that are based on a 
range of different values for $\alpha_{s}$. 
The purpose of this paper is to fill that need.

We provide here a series of PDF sets that span a range 
of coupling strengths from $\alpha_{s}(m_{Z}) = 0.110$ to $0.128$.
These PDFs extend and update the CTEQ6 global analysis \cite{cteq6}.
They are successors to the $\alpha_{s}$ series that was 
determined in the CTEQ4 analysis \cite{cteq4}.
Complementary to the physics probed in \cite{cteq6},
we will discuss several issues related to the dependence
on $\alpha_{s}$:
the correlation between $\alpha_{s}$ and the parton 
distributions---particularly the gluon distribution;
the viability of using global analysis of hadronic processes
to measure $\alpha_{s}$;
and the dependence on $\alpha_{s}$ of physics predictions
for $W$, $Z$, jet, and Higgs boson cross sections
at the Tevatron and the Large Hadron Collider LHC.

We provide two different sets of 
fits---CTEQ6A and CTEQ6B---corresponding to two different 
common definitions for $\alpha_s(\mu)$ that are defined 
in the next Section.

\section{Definitions for $\protect{\mathbf{\alpha_s(\protect\mu)}}$}

\label{sec:alphas}

The dependence of the QCD coupling strength $\alpha_{s}(\mu )$
on the momentum scale $\mu $ is governed by the renormalization
group equation (RGE), which in
next-to-leading order (NLO) perturbation theory is
\begin{equation}
\mu \,d\alpha /d\mu =c_{1}\alpha^{2}\,+\,c_{2}\alpha^{3}\;,
\label{RGE}
\end{equation}
where $c_{1}=-\beta_{0}/2\pi $ with $\beta_{0}=11-(2/3)\mathrm{n}_{f}$,
and $c_{2}=-\beta_{1}/8\pi ^{2}$ with
$\beta_{1}=102-(38/3)\mathrm{n}_{f}$.
The coefficients are functions of the number of active
quark flavors $\mathrm{n}_{f}$. 
These coefficients change discontinuously as $\mu$ passes the mass 
$m_{f}$ of each quark flavor and the integer $\mathrm{n}_{f}$ jumps 
up by $1$; nevertheless at NLO,
the function $\alpha_{s}(\mu )$ is a \textit{continuous} function
of $\mu$ at these thresholds.
The solution to Eq.~(\ref{RGE}) therefore depends on a single integration 
constant, which is the fundamental coupling strength parameter of QCD 
that is usually chosen to be $\alpha_s(m_Z)$.

Since Eq.~(\ref{RGE}) is truncated at ${\cal O}(\alpha^{3})$, there are 
infinitely many definitions of $\alpha_{s}(\mu )$ that are formally
equivalent because they differ only in higher orders of the perturbative 
expansion.
Two choices that are often used in QCD phenomenology are:
\begin{itemize}
\item{
{\bf Def.~A.} The original NLO definition \cite{MsbAlf} is given by
\begin{equation}
\alpha_{s}(\mu )=c_{3}\,[1-c_{4}\ln (L)/L]/L\;,
\label{MsbarAlfa}
\end{equation}
where $L=\ln (\mu ^{2}/\Lambda ^{2})$, $c_{3}=-2/c_{1}$, and 
$c_{4}=-2\,c_{2}/c_{1}^{\,2}\,$.
The parameter $\Lambda$ depends on $\mathrm{n}_{f}$, and hence takes on 
different values $\Lambda_{n_f}$ when $\mu$ 
crosses each quark mass threshold.
Previous CTEQ global analyses have all used this definition
for the running coupling.
}
\item{
{\bf Def.~B.} An alternative is to solve the truncated RGE (\ref{RGE}) 
exactly.  The publicly available evolution program QCDNUM \cite{QCDNUM},
and many HERA analyses use this definition.
}
\end{itemize}

We have shown previously \cite{stability} that these two forms are
numerically quite similar in the region $Q>2\,\mathrm{GeV}$
where we fit data.
In the following, we present results based on both 
definitions,\footnote{The form used for PDF fitting by 
MRST \cite{mrst2002} is different from both of these definitions, 
but is numerically very close to Def.~B \cite{stability}.  
The definition used by the Particle Data Group \cite{PDG} lies between 
Def.~A and Def.~B.}

\section{The CTEQ6 $\protect{\mathbf{\alpha_{s}}}$ series of global fits}

\label{sec:series}

To study the interplay between the strong coupling strength $\alpha_{s}$
and the PDFs, one can either perform a fully global QCD fit by varying 
$\alpha_{s}$ and the PDF parameters simultaneously to 
examine the neighborhood of the overall global minimum in $\chi^{2}$; 
or one can perform a series of fits to the
PDF parameters at various fixed values for $\alpha_{s}$.
We have explored both of these approaches, but concentrate here on the 
second approach since it is most convenient for general collider physics 
applications.  
We therefore present here a series of 
PDFs---the CTEQ6 $\alpha_{s}$-series.
Some of their physical implications are discussed in 
Sec.\ \ref{sec:predictions}.
Earlier PDF fits with a range of $\alpha_{s}(m_{Z})$ were
obtained in \cite{cteq4,MRSTseries}.  Fits with a range of $\alpha_{s}(m_{Z})$
and their interplay with the gluon distribution were studied in connection with
the possibility of coevolution with light gluinos in \cite{gluino}.

The PDFs in the new series were obtained for 10 values of
$\alpha_{s}(m_{Z})=0.110$, $0.112$, \dots, $0.128$,
where $m_Z = 91.188 \, \mathrm{GeV}$ is assumed.
The theoretical assumptions and functional parametrization
of these PDFs at the initial momentum scale 
$\mu_{0}=1.3 \, \mathrm{GeV}$
are the same as in the previous CTEQ6
analysis \cite{cteq6,c6jet}. 
The experimental input is slightly updated.\footnote{%
The input experimental data consist of 
BCDMS (muon $F_2$ on hydrogen and deuterium) \cite{bcdms}, 
H1 ($e^{\pm}$ $F_2$) \cite{H1},
ZEUS ($e^{\pm}$ $F_2$) \cite{zeus}, 
NMC (muon $F_2$ on hydrogen and deuterium) \cite{nmc}, 
CCFR (neutrino $F_2$ and $xF_3$ on iron) \cite{ccfr}, 
E866 (Drell-Yan muon pairs from $pp$ and $pd$) \cite{e866},
CDF (W-lepton asymmetry) \cite{cdfWasym}, 
CDF (inclusive jet) \cite{cdfJet} and 
D0 (inclusive jet) \cite{D0Jet}. 
The update consists mainly of including the data from the 
first of the three references in \cite{H1} which was 
inadvertently omitted in CTEQ6.}
These PDF sets are designated as CTEQ6A110, \dots, CTEQ6A128
using Def.~A for $\alpha_{s}$;
and CTEQ6B110, \dots, CTEQ6B128 using Def.~B for $\alpha_{s}$.
The fits
CTEQ6A118 and CTEQ6B118, which have $\alpha_{s}(m_{Z})=0.118$,
are nearly the same as CTEQ6M or CTEQ6.1M,
but are not identical to either due to minor 
updates in experimental input, and in the case of CTEQ6B118 the different 
definition for $\alpha_{s}$.
All 20 of these PDF sets will give very similar
physical predictions for most applications.

Figure \ref{fig:figalphaseries1x} shows the quality of the global fits,
as measured by the overall $\chi^{2}$ for the fit
to $\sim \! 2000$ data points,
as a function of $\alpha_{s}(m_{Z})$.\footnote{%
This figure is an improved version of one appearing in \cite{stability}.}
The two curves, both approximately parabolic, are
smooth interpolations of the above series of fits. 
The curves are very similar, with the Def.~A curve being 
slightly narrower and slightly farther to the left because 
Def.~A has a little more rapid variation of $\alpha_{s}(\mu)$ with $\mu$.
The minima of these curves are at $\alpha_{s}(m_Z) = 0.1172$ and $0.1176$, 
close to the current world average $(0.1187\pm 0.0020)$ \cite{PDG}
and very close to the value $0.1180$ used in CTEQ6M and CTEQ6.1M.
This similarity in $\alpha_{s}$ is an impressive demonstration
of consistency between QCD theory and experiment,
since the global QCD analysis is based on hard scattering
data with hadronic initial states,
while the determination of the world average value for $\alpha_{s}$
comes mainly from totally different physical processes such as
$e^{+}e^{-}$ annihilation, $\tau $-decay,
and even lattice gauge theory calculations with quarkonium spectra as input.

Note that the range of $\alpha_{s}(m_{Z})$ covered by the
CTEQ6AB series is much wider than the currently
accepted $1\sigma$ error range of the world average quoted above.
We explore this extended range because 
the lowest and highest values ($0.110$, $0.128$) represent outlying values 
that have been obtained by some individual experiments that are included in 
the world average.
The fits with extreme values of $\alpha_{s}(m_{Z})$ will be useful for 
some specialized applications.
In plots shown later in this paper, we show almost all of this range, 
but reduce it to $0.110$ to $0.126$ to be symmetric about the CTEQ6
value of $0.118$.
Note that $\chi^{2}$ increases by $\sim \! 100$ above its minimum
at the extremes of this reduced range, which makes it consistent 
with a $90\%$ confidence range for the global fit according to 
results of our previous analyses.
However, the reader must keep in mind that this range of $\alpha_{s}(m_{Z})$ 
is larger by a factor of $\sim \!4$ than the uncertainty range corresponding
to a ``$1\sigma$'' error band based on the world average data.

%

\section{The gluon distribution and $\protect{\mathbf{\alpha_{s}}}$}

\label{sec:gluons}

The gluon distribution is strongly correlated with $\alpha_{s}$ in the 
global QCD analysis.  This can be seen in Fig.~\ref{fig:figquark10c12}(a), 
which shows the gluon distributions $g(x,\mu)$ from the $\alpha_{s}$-series 
PDFs as a function of momentum fraction $x$ at scale 
$\mu=3.162 \, \mathrm{GeV}$ ($\mu^2 = 10 \, \mathrm{GeV^2}$).  
For clarity of display, 
the horizontal axis is scaled as $x^{1/3}$ and the vertical axis is 
weighted by $x^{3/2}$.

For comparison, Fig.~\ref{fig:figquark10c12}(b) shows the uncertainty 
band (shaded area) of the gluon distribution due to sources other than 
$\alpha_{s}$.  This uncertainty was computed at a fixed value 
$\alpha_{s}(m_{Z}) = 0.118$ by the Hessian method \cite{40sets}, using 
the 40 eigenvector basis sets of CTEQ6.1 \cite{c6jet}.
The CTEQ6.1M (solid) and the new CTEQ6A118 (dashed) distributions, which 
are very similar, are also shown in  Fig.~\ref{fig:figquark10c12}(b).
The two figures are combined in  Fig.~\ref{fig:figquark10c12}(c),
where, in order to highlight the differences, the results are shown as 
ratios to the CTEQ6.1M distribution.

There is a clear systematic trend in the $\alpha_{s}$-series
for the gluon distribution function.
Fits with larger $\alpha_{s}(m_{Z})$ have a gluon component that is
weaker at small $x$ and stronger at large $x$.
The behavior at small $x$ results from the fact that 
every occurrence of $g(x,\mu )$ in a cross section formula
is accompanied by a factor of $\alpha_{s}$,
so when  $\alpha_{s}$ is made larger, the gluon distribution 
becomes smaller in order to maintain agreement with the large 
amount of data at small $x$.
The behavior at large $x$, where the direct experimental constraints 
on the gluon are weaker, is dictated by the momentum sum 
rule: the total momentum fraction carried by gluon + quarks must be 
equal to 1, and the momentum carried by quarks is tightly constrained 
by DIS data.

Another feature seen in Fig.~\ref{fig:figquark10c12} is that 
the gluon distributions for different $\alpha_{s}(m_{Z})$ values all
nearly intersect at a common value $x \approx 0.1$.
This occurs simply because the function $g(x, \mu, \alpha_{s}(m_{Z}))$
varies rather slowly and smoothly with $\alpha_{s}(m_{Z}))$, so its
dependence on $\alpha_{s}(m_{Z}))$ can be approximated rather well 
by the first order (linear) term of a Taylor series.  This linearity 
can also be seen by the nearly equal spacing of the curves in 
Fig.~\ref{fig:figquark10c12}(a) and (c).

Comparing the range of variation of $g(x,\mu)$ due to the variation of
$\alpha_{s}$, to the uncertainty range due to other sources of error,
we see that the two are comparable throughout most of the domain in $x$.
The {\em combined} uncertainty on the gluon distribution is therefore
somewhat larger than the previously published uncertainties,
which were obtained at fixed $\alpha_{s}(m_{Z})$.

Figure~\ref{fig:figquark13de}(a) shows the gluon uncertainty 
at $\mu = 100 \, \mathrm{GeV}$.  At this larger momentum scale, 
the overall uncertainty is much smaller, and the $\alpha_{s}$ 
contribution to the uncertainty is generally smaller than the 
uncertainty due to the causes that are included in the Hessian analysis.
Figure~\ref{fig:figquark13de}(b) similarly shows the uncertainty for the
up quark distribution at $\mu = 100 \, \mathrm{GeV}$.  The uncertainty
for the quark is much smaller than for gluon (note the different 
y-axis in the graph) and again the $\alpha_{s}$ contribution is 
small compared to the other sources of uncertainty.

\section{Can we determine $\protect{\mathbf{\alpha_{s}}}$ from the 
global analysis?}

\label{sec:DetermineAlpha}

It is natural to try to determine the coupling strength $\alpha_{s}$ 
from the global QCD fit to hadronic processes.  It appears straightforward 
to do so, since one can simply treat the parameter
$\alpha_{s}(m_{Z})$ as one of the fitting parameters.
We see from the minima in Fig.\ \ref{fig:figalphaseries1x}
that the resulting ``best fit'' value of $\alpha_{s}(m_{Z})$ is 
around $0.1174$, which is very close to the world average value.
The difficult question, however, is
what uncertainty should be assigned to this measurement.
The answer to that question determines whether this method
is competitive with measurements that are independent of the
complications of hadron partonic structure.
The fact that the value of $\alpha_{s}(m_{Z})$ in global analysis is
strongly correlated with the rather uncertain gluon distribution, as
discussed in the previous section, suggests that caution is needed.

Referring again to Fig.\ \ref{fig:figalphaseries1x},
the range of $\alpha_{s}(m_{Z})$ (horizontal axis) allowed
by global analysis based on minimization of $\chi^{2}$,
is set by the increase of $\chi^{2}$ (vertical axis)
above the global minimum that we allow.
We refer to the allowed increase of $\chi^{2}$ as the {\em tolerance},
$\Delta\chi^{2}$.
Recent studies of uncertainties of PDFs by various global analysis
groups \cite{cteq6,mrst2002,zeusglobalfit} have concluded that
a reasonable tolerance $\Delta\chi^{2}$ must be rather large,
in the range $50$ -- $100$ for the $\sim \! 2000$ points in 
present-day data sets, to define an approximate $90\%$ confidence range.
Making the specific choice $\Delta\chi^{2}=100$, the corresponding
range of $\alpha_{s}(m_{Z})$ is from 
$0.1093$ to $0.1247$.
Assuming that range to be the $90\%$ confidence range
for a gaussian distribution,
it corresponds to $\alpha_{s}(m_{Z}) = 0.1170 \pm 0.0047$
for a ``$1\sigma$'' error range.
Thus the measurement of $\alpha_{s}(m_{Z})$ provided by 
PDF fitting agrees very well with the 
Particle Data Group world average of $0.1187\pm 0.0020$ \cite{PDG},
but has more than twice its uncertainty.  Hence the PDF result cannot be 
used to substantially reduce the uncertainty in $\alpha_{s}(m_{Z})$ at the 
present time. 

We can gain some insight on how $\alpha_{s}$ is constrained in the global
QCD analysis by examining the dependence on $\alpha_{s}(m_{Z})$
of the $\chi^{2}$ values for each individual experiment that contributes
to that analysis.
These $\chi^{2}$ ``parabolas'' are shown in Fig.\ \ref{fig:ParabExpts}.
The curves represent smooth interpolations of the results from
the 10 fits in the CTEQA $\alpha_{s}$-series.
The vertical axis in each graph is the $\chi^{2}$ value
per data point in the fit to that experiment, while the 
horizontal axis is $\alpha_{s}(m_{z})$.

It is apparent in Fig.\ \ref{fig:ParabExpts}
that the sensitivities of the various experiments
to $\alpha_{s}$ vary greatly.
Some of the curves are approximately parabolic with a minimum
within the range probed, while
others merely constrain the value of $\alpha_{s}(m_{Z})$
from above or below.
The global minimum seen in Fig.\ \ref{fig:figalphaseries1x}
is due to the combined constraints of all the experiments.
Since different experiments prefer different values
of $\alpha_{s}(m_{Z})$,
which are not always consistent with each other if strict
statistical criteria (``$\Delta \chi^2 = 1$'') are applied to 
each experiment, the global minimum represents a compromise that 
is difficult to interpret
as a ``measurement'' in the traditional sense.\footnote{%
We should point out that the $\chi^{2}$ curves shown in
Fig.\ \ref{fig:ParabExpts} are not to be compared directly to
those obtained by individual experiments in their respective
determinations of $\alpha_{s}$.
The points on our curves correspond to $\chi^{2}$ values evaluated using 
constrained fits to the full global data set,
not just to the data of a single experiment.
}
In particular, there is no clear way to assign
a statistically meaningful error to the measurement.
Rather, the error is dominated by systematic effects that 
can only be estimated.


\section{Dependence of predicted cross sections on $\protect{\mathbf{\alpha_{s}}}$} 
\label{sec:predictions}

Here we present predictions for several important processes at the 
Tevatron ($\bar{\mathrm{p}} \mathrm{p}$ at 
$\sqrt{s} = 1.96 \, \mathrm{TeV}$) and 
the LHC ($\mathrm{pp}$ at $\sqrt{s} = 14 \, \mathrm{TeV}$). 
The results show how the uncertainty in $\alpha_{s}$ propagates
to uncertainties in physical predictions.
At the same time, the results show to what extent accurate measurements 
of these cross sections could be used to constrain $\alpha_{s}$.

We show predictions using the Def.~A form for $\alpha_{s}$.
Results for the Def.~B form are very similar.
The figures show the variation of the predictions for the 
range of $\alpha_{s}(m_{Z})$ from 0.110 to 0.126.
We again remind the reader that this range is much larger
than the actual uncertainty of $\alpha_{s}(m_{Z})$;
i.e., the full variation of the physical predictions shown in each
figure extends beyond the actual uncertainty from $\alpha_{s}$.
The shaded region in each figure shows the range of uncertainty
due to sources other than $\alpha_{s}$, as calculated 
from the eigenvector basis sets
of CTEQ6.1 using the Hessian method \cite{c6jet,40sets}.  
These are assumed to estimate the $90\%$ confidence range.

\subsection{$W$ and $Z$ production}

\label{sec:WZ}

Figure \ref{fig:figWZd} shows the cross section for $W^{-}$ production
at the Tevatron.\footnote{All cross sections shown for $W^{\pm}$ and $Z^0$ 
production are given in nanobarns, with the leptonic decay branching 
fraction included.}
$W^{+}$ production is identical except for $y \to -y$.
The left plot shows $d\sigma/dy_{W}$ versus the $W$ rapidity $y_{W}$.
The right plot shows the difference from the prediction of 
CTEQ6.1, which has $\alpha_{s}(m_{Z})=0.118$.
The curves are the CTEQ6A $\alpha_{s}$-series, and  
the shaded band is the range of uncertainty for fixed $\alpha_{s}(m_{Z})$,
calculated using the Hessian method.
For this process, the Hessian uncertainty range is about
$\pm 5\%$ of the central prediction.
The variation with $\alpha_{s}$ is smaller, on the order of $\pm 2\%$,
even for the extreme range of $\alpha_{s}(m_{Z})$ from 
$0.110$ to $0.126$.

Figure \ref{fig:figWZf} shows the cross section $d\sigma/dy_{Z}$ for $Z^{0}$
production at the Tevatron.
Again the Hessian uncertainty range is $\sim \pm 5\%$ and again
the variation with $\alpha_{s}(m_{Z})$ is 
$\sim \pm 2\%$ for the extreme range of $\alpha_{s}$.

Figure \ref{fig:figWZa} shows the cross section for $W^{-}$ production 
at the LHC.  
The Hessian uncertainty range is again of order $\pm 5\%$,
but the variation with $\alpha_{s}$ is larger than for
the Tevatron, of order $\pm 5\%$ for the large range of $\alpha_{s}(m_{Z})$
that is shown.
Figure \ref{fig:figWZb} shows the process of $W^{+}$ production at the LHC.
It has a larger cross section at large rapidity (because $u(x) > d(x)$ for 
the valence quarks) and a similar range of uncertainty in the prediction.
The difference between the central dashed curve and a horizontal line in 
Figs.~\ref{fig:figWZa} and \ref{fig:figWZb} shows the effect of updates in 
the fitting between CTEQ6.1 and CTEQ6A118: the change is well within the 
estimated PDF errors.

The cross sections for $W^\pm$ production at the LHC are closely tied
to the gluon distribution, since the leading-order process in 
proton-proton collisions is $u+\overline{d}\rightarrow W^{+}$, 
which involves a sea quark; and sea quarks and the gluon are 
related at large $\mu$ by the evolution equations.
(Proton-antiproton collisions at the Tevatron are different
because both $u$ and $\overline{d}$ can be valence quarks---which 
causes the asymmetry in $y$ that can be seen in Fig.\ \ref{fig:figWZd}.)
Also, the next-to-leading order process
$u + g \rightarrow d + W^{+}$ involves an initial gluon directly.
Hence the predictions for $W$ production at the LHC are somewhat 
more sensitive to $\alpha_{s}$ than are the predictions for the 
Tevatron.  Nevertheless, the uncertainty associated with $\alpha_{s}$
at the LHC, for values of $\alpha_{s}$ that are consistent with the 
world average, remains small compared to the other PDF uncertainties 
as can be seen in Figs.~\ref{fig:figWZa}(b) and \ref{fig:figWZb}(b).

\subsection{Inclusive Jets}
\label{sec:jets}

Figure~\ref{fig:figjet1} shows the $\alpha_{s}$-dependence of 
predictions for inclusive jet production at the Tevatron
(in the CDF central rapidity region $0.1 < |y| < 0.7$), and 
at the LHC (in the region $|y| < 1$).
The $\alpha_{s}$ dependence is exhibited by plotting the ratio of the
predicted cross section $d\sigma/dp_{T}$ to the prediction calculated 
using CTEQ6.1.
As before, the shaded region is the uncertainty range for fixed 
$\alpha_{s}(m_{Z}) = 0.118$ calculated from the eigenvector basis sets 
of the Hessian approach \cite{c6jet,40sets} and the curves show 
$\alpha_{s}(m_{Z}) = 0.110$, \dots, $0.126$.

We observe that if we restrict $\alpha_{s}(m_{Z})$ to a range 
$\pm 0.003$ which is the $90\%$ confidence range of the world 
average, the uncertainty due to $\alpha_{s}(m_{Z})$ is small 
compared to the other PDF uncertainties at large $p_T$.
However, at moderately small $p_T$ the uncertainty in 
$\alpha_{s}(m_{Z})$ adds considerably to the overall uncertainty 
for jet production.
This comes about because the parton distributions do not depend 
very strongly on $\alpha_{s}(m_{Z})$ 
(see Fig.~\ref{fig:figquark13de}), but the hard cross sections do.

\subsection{Higgs boson cross section in SM and MSSM}

The uncertainty of the cross section predicted for Higgs boson
production at the LHC as a function of Higgs mass is shown in
Fig.~\ref{fig:figHiggsBoth}.
Figure~\ref{fig:figHiggsBoth}(a) shows the uncertainty 
for the $gg \to H$ process,  while
Fig.~\ref{fig:figHiggsBoth}(b) shows the uncertainty 
for the $b \bar{b} \to H$ mechanism. These cross sections
were calculated at NLO using programs from 
\cite{Spira:1995rr,Spira:1996if} and 
\cite{Balazs:1998sb} respectively.
Both processes play an important role in Higgs physics:
the $gg$ process is dominant in the Standard Model (SM),
while in Supersymmetry and some other generic extensions
of SM, the $b\bar{b}\to H$ process can be equally important 
or even dominant.
The study of PDF uncertainties is therefore important
for both mechanisms of Higgs production.

The uncertainties based on the 40 eigenvector sets of CTEQ6.1 are 
shown as the shaded regions.  These have been shown previously for 
$gg \to H$ \cite{Djoudi} and for $b \bar{b} \to H$ \cite{Belayev}.
The 9 curves in each figure show the predictions of the PDF sets 
with $\alpha_s(m_Z) = 0.110, \dots, 0.126$
in steps of $0.002$ relative to the central value $0.118$, which
therefore corresponds to the horizontal line at ratio $1.0$.

Figure~\ref{fig:figHiggsBoth}(a) shows that the uncertainty in 
$\alpha_s$ substantially increases the uncertainty in the prediction 
for $g g \to H$.  Particularly for $M_H\simeq 300 \, \mathrm{GeV}$
where the non-$\alpha_s$ PDF uncertainty is a minimum, the uncertainty 
due to $\alpha_s$ is larger than the non-$\alpha_s$ uncertainty.
The strong sensitivity of $g g \to H$ to $\alpha_s$ is of course not 
surprising in view of the $\alpha_s^2$ dependence coming from the 
leading order triangle diagram; indeed because the K-factor is large 
and positive for this process, the NLO corrections make the dependence 
on $\alpha_s$ even stronger.\footnote{This simple notion agrees quite 
well quantitatively with the results in Fig.~\ref{fig:figHiggsBoth}(a).  
For instance, at $M_H = 100 \, \mathrm{GeV}$, raising $\alpha_s$ by 
$5\%$ from $0.118$ to $0.124$ would be expected to raise the cross 
section by about $13\%$---halfway between 
$10\%$ (for $\alpha_s^2$) and 
$16\%$ (for $\alpha_s^3$).  But meanwhile, $M_H = 100 \, \mathrm{GeV}$ 
requires $x \approx 0.007$, where the increase in $\alpha_s$ causes 
$g(x)$ to decrease by about $3\%$ according to 
Fig.~\ref{fig:figquark13de}(a), which would lower the cross section 
for $g g \to H$ by $6\%$.  Combining these two effects, the net change 
is an increase of $7\%$, which agrees well with the actual increase 
of $8\%$ that is seen in Fig.~\ref{fig:figHiggsBoth}(a).
}

Figure~\ref{fig:figHiggsBoth}(b) shows that the contribution of 
the uncertainty in $\alpha_s$ to $b\bar{b}\to H$ production is 
comparable to the other PDF uncertainties for that process, so its 
effect is to add just a modest increase in the overall uncertainty 
for that process.  This is not surprising, since $b\bar{b}\to H$ 
has no direct dependence on $\alpha_s$ at leading order, and the 
quark distributions at this momentum scale do not vary rapidly 
with $\alpha_s$ as seen in Fig.~\ref{fig:figHiggsBoth}(b).

Figure~\ref{fig:figHiggsPct} summarizes the results of 
Fig.~\ref{fig:figHiggsBoth}:  The dotted curves are the PDF 
uncertainty at fixed $\alpha_s$ (identical to the 
boundary of the shaded regions in Fig.~\ref{fig:figHiggsBoth})
which are intended to show a $90\%$ confidence range.
The dashed curve is the uncertainty due to $\alpha_s$, 
calculated by interpolation for an uncertainty of 
$\pm 1.64 \, \sigma$ in $\alpha_s(m_Z)$, where $\sigma=0.002$
from the world average and the factor $1.64$ corresponds to 
a $90\%$ confidence range.  The solid curve shows the 
combined uncertainty obtained by adding the two contributions 
in quadrature.\footnote{%
Adding the errors in quadrature, i.e., treating the 
additional source of error due to $\alpha_s(m_Z)$ as independent 
of the other PDF errors, is 
the correct approach according to the Hessian approximation.
For in that approximation, $\chi^2 = \chi_0^2 + \sum_i z_i^{\,2}$
where the $z_i$ are linear combinations of the PDF shape 
parameters $A_i$: $z_i = \sum_j T_{ij} A_j$.  When a new 
parameter such as $(\alpha_s(m_Z) - 0.118)$ is added to the set 
of fitting parameters, $\chi_0^2$ becomes a quadratic function 
of that parameter, but there is no change in $T_{ij}$ because 
in the Hessian approximation one drops all contributions to 
$\chi^2$ that are higher order than quadratic.
}
The additional uncertainty due to $\alpha_s$ is seen to be 
substantial for the $gg \to H$ process.

\section{Conclusion}

Previous CTEQ6 parton distributions were extracted from 
experiment assuming $\alpha_{s}(m_{Z}) = 0.118$, based on the 
world average value.
The PDFs presented here were extracted using a range of 
alternative assumptions for $\alpha_{s}(m_{Z})$, using a 
similar set of experiments with only minor updates. 

The new PDFs are named 
CTEQ6A110, \dots, CTEQ6A128, and 
CTEQ6B110, \dots, CTEQ6B128, where the ``A'' or ``B'' label indicates 
the choice 
of functional form for $\alpha_s(\mu)$ (see Sec.~\ref{sec:alphas}) 
and ``110'' e.g. indicates $\alpha_s(91.188 \, \mathrm{GeV}) = 0.110$.
Fortran programs to calculate these PDFs are available at 
the LHAPDF archive http://www-spires.dur.ac.uk/HEPDATA/.

These PDFs can be used to find the range of uncertainty in predictions 
due to the uncertainty in $\alpha_{s}(m_{Z})$.  We find that the previous 
fits with $\alpha_{s}(m_{Z}) = 0.118$ are adequate for most processes,
because the uncertainty associated with $\alpha_{s}(m_{Z})$ is smaller 
than the other sources of PDF uncertainty.  However, the 
$\alpha_{s}(m_{Z})$ uncertainty is important for some predictions that 
are particularly sensitive to $\alpha_s$ or to the gluon distribution, 
such as inclusive jet production at relatively small $p_T$ 
(see Fig.~\ref{fig:figjet1}) and Higgs boson production by the 
$gg \to H$ process in the standard model 
(see Fig.~\ref{fig:figHiggsPct}(a)).  This comes about because the 
intrinsic $\alpha_{s}^2$ that is present in the hard scattering processes
is only partly compensated by changes in the PDFs with $\alpha_{s}(m_{Z})$
such as those shown in Fig.~\ref{fig:figquark13de}.

These PDFs can also be used to study the constraints from global fitting 
or to study individual experiments on the value of $\alpha_{s}(m_{Z})$.
In the latter case, the new alpha-series PDFs allow one to take into 
account the strong correlation between $\alpha_{s}(m_{Z})$ and the gluon 
distribution that is present at small momentum scales, as shown in 
Fig.~\ref{fig:figquark10c12}.

\paragraph{Acknowledgements:}
We thank C.-P. Yuan useful discussions.
This research is supported by the National Science Foundation.

\figalphaseriesonex
\figquarktenconetwo
\figquarkde
\figParabExpts
\figWZd 
\figWZf 
\figWZa 
\figWZb 
\figjetone
\figHiggsBoth
\figHiggsPct

\end{document}